\documentclass[conference, final]{IEEEtran}
\IEEEoverridecommandlockouts

\usepackage[noadjust]{cite}
\usepackage{amsmath,amssymb,amsfonts}
\usepackage{algorithmic}
\usepackage{graphicx}
\usepackage{textcomp}
\usepackage{xcolor}
\usepackage{booktabs}
\usepackage{tabularx}
\usepackage{comment}
\def\BibTeX{{\rm B\kern-.05em{\sc i\kern-.025em b}\kern-.08em
    T\kern-.1667em\lower.7ex\hbox{E}\kern-.125emX}}

\IEEEoverridecommandlockouts
\IEEEpubid{\makebox[\columnwidth]{979-8-3315-5021-9/26/\$31.00~\copyright2026~IEEE \hfill} \hspace{\columnsep}\makebox[\columnwidth]{ }}
\begin{document}

\title{RelicTouch: Exploring Fingertip Vibrotactile Feedback for Image Interpretation}

\author{\IEEEauthorblockN{Xiaosong Wang\IEEEauthorrefmark{1}%
\thanks{\IEEEauthorrefmark{1}Corresponding author.}}
\IEEEauthorblockA{\textit{School of Art and Archaeology} \\
\textit{Zhejiang University}\\
Hangzhou, China \\
wang64@zju.edu.cn}
\and
\IEEEauthorblockN{Ran Liu}
\IEEEauthorblockA{\textit{College of Computer Science and Technology} \\
\textit{Zhejiang University}\\
Hangzhou, China \\
liuran@zju.edu.cn}
\and
\IEEEauthorblockN{Yifan Feng\IEEEauthorrefmark{2}%
\thanks{\IEEEauthorrefmark{2}Also with Creative Computing Institute, University of the Arts London.}}
\IEEEauthorblockA{\textit{UCL Interaction Center} \\
\textit{University College London}\\
London, United Kingdom \\
yifanfeng4@acm.org}
}

\maketitle

\begin{abstract}
Vibrotactile feedback is often used to simulate
material properties on interactive surfaces, but less is known
about how it can help users interpret visual differences during
touchscreen exploration. We present \textit{RelicTouch}, a
finger-worn system that converts image-derived features into
movement-contingent vibration. Study~1 was a controlled
within-subject experiment ($n=20$) examining how vibration
intensity, stripe spacing, and finger configuration affected
haptic--visual congruence and subjective clarity. Study~2 was
an exploratory qualitative study with eight domain experts examining how users
interpreted cues generated from luminance and edge variation
during continuous image exploration. Results suggest that cue
legibility was associated with distinguishable signal changes,
movement-based comparison, and visual or task context.
Single-finger exploration yielded higher subjective ratings,
whereas multi-finger feedback showed possible value for coarse
comparison across separated features. These findings provide
initial design implications for vibrotactile image exploration
beyond material-simulation-oriented rendering.


\end{abstract}


\section{Introduction} 
Touch provides a fundamental means for perceiving structure through active exploration. While modern touchscreens enable rich visual interaction, they lack mechanisms for conveying spatial structure through touch. Prior haptic research has therefore focused on augmenting tactile feedback on planar interfaces, most commonly by simulating material properties such as roughness or compliance~\cite{ding2017tactile,asano2012vibrotactile,ito2019tactile,vega2024varitouch}. Although these approaches enhance local tactile sensations, they implicitly assume that haptic feedback should approximate physical materials, and are therefore limited in supporting the interpretation of complex, image-based content.

Prior HCI and haptics research has demonstrated that vibrotactile feedback can simulate textures or encode spatial features, particularly in accessibility and non-visual interaction contexts~\cite{hwang2025computational}, but less is known about how users interpret such encoded signals when they actively explore image-based content on touchscreens. In particular, it remains unclear how signal parameters, finger configuration, and movement strategies jointly shape whether vibrotactile cues become legible as spatial structure.
To address this gap, we position fingertip vibrotactile feedback not as a means of reproducing material properties, but as an interpretive encoding of spatial structure that unfolds through interaction. Rather than assuming that tactile signals carry fixed meaning, we investigate how meaning emerges as users explore, compare, and interpret variations in vibration across space. This perspective treats vibrotactile parameters as perceptual
constraints on interpretation rather than as indicators of
fidelity to physical materials.


\begin{figure}[!bt]
    \centering
    \includegraphics[width=\linewidth]{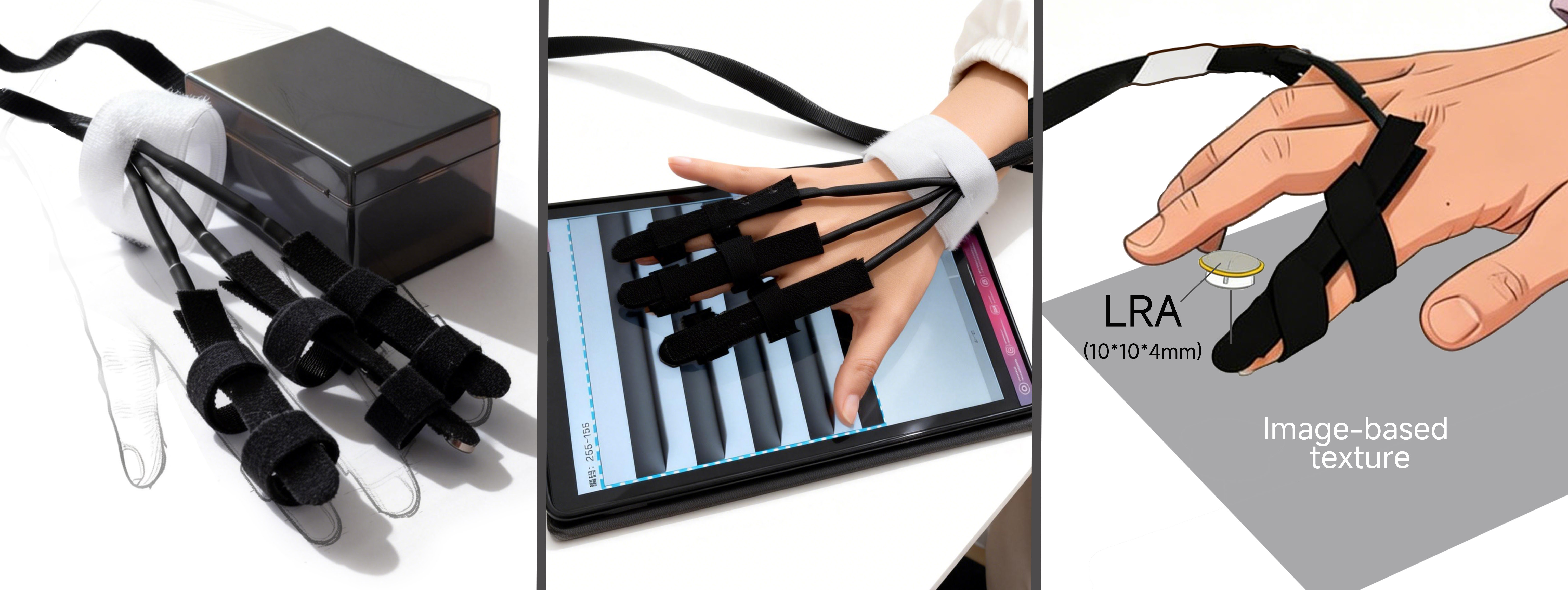}
    \caption{RelicTouch: a fingertip haptic wearable system}
    \vspace{-16pt}
    \label{fig:system4}
    
\end{figure}

We introduce \textit{RelicTouch}, a fingertip-worn vibrotactile system that enables users to explore image-based content on touchscreens through movement-contingent tactile feedback. As users move their fingers across an image, vibration is dynamically modulated based on image-derived features, allowing spatial variations to be perceived through touch. We use \textit{RelicTouch} as an interaction technique and a probe to investigate how vibrotactile signals are interpreted during continuous exploration and conduct a two-stage experiment as experience evaluation.  Study 1 first establishes the perceptual operating conditions for RelicTouch through a controlled within-subject experiment. It examines how vibration intensity, stripe spacing, and finger configuration influence haptic--visual congruence and subjective clarity, identifying usable regions within the intensity--spacing parameter space. Study 2 then applies these findings to a continuous image-exploration context. Rather than repeating the full parameter comparison, Study 2 uses the perceptually constrained intensity range and single-finger condition identified from Study 1 to examine how participants interpret image-derived vibrotactile cues during open-ended exploration. In summary, this paper makes \textbf{three contributions}:
First, we present \textit{RelicTouch}, a fingertip-worn vibrotactile system that encodes image-derived spatial variation as movement-contingent tactile feedback during touchscreen exploration; Second, we report a controlled within-subject study showing how vibration intensity, stripe spacing, and finger configuration affect haptic--visual congruence and subjective clarity, characterizing more frequently selected regions within the tested intensity--spacing parameter space; Third, we derive three situated design implications for vibrotactile image exploration, grounded in findings from both studies.  These implications concern signal separability, movement-contingent comparison, and matching feedback channels to the spatial scale of image content.


\section{Related work}

\subsection{Vibrotactile Rendering for Texture and Spatial Perception}
Vibrotactile feedback has been widely used to render material-related tactile qualities during touch interaction. Prior work has used vibrotactile feedback to render material qualities such as roughness and compliance during sliding or pressing interactions ~\cite{Ujitoko2019, Baeck2025, Preechayasomboon2021,sabnis2025designing,strohmeier2020barefoot,heo2019pseudobend,vega2024varitouch,sabnis2023haptic}. These studies typically establish mappings between vibration parameters and perceived tactile qualities, aiming to approximate real-world surface sensations. At the same time, accessibility and non-visual interaction research has shown that vibrotactile cues can encode spatial information, such as shapes, edges, or layouts~\cite{hwang2025computational, jiang2024designing, paneels2009review}. These works demonstrate that tactile feedback can convey abstract spatial information beyond material properties. 

\subsection{Parameter-Driven Vibrotactile Encoding}
Much research has examined how vibrotactile parameters such as intensity, frequency, and spatial resolution affect perception, establishing important constraints including detection thresholds, spatial acuity, and discrimination limits ~\cite{chen2018research, choi2020augmenting,jingu2024shaping, jiang2024designing, palani2018touchscreen}. For example, studies on vibrotactile spatial acuity have investigated how intensity and frequency influence the resolution of tactile displays~\cite{huang2024investigating,chen2025low}, while other work has evaluated users' ability to distinguish vibrotactile patterns under controlled conditions \cite{vega2024varitouch}. These studies show that vibrotactile perception is bounded by signal properties and perceptual thresholds. However, most focus on isolated stimuli or discrete discrimination tasks. In contrast, image-based exploration requires users to interpret continuously changing vibration patterns as they move across visual content. We therefore treat vibrotactile parameters not only as controllable inputs, but also as perceptual constraints that shape the legibility of tactile cues during active exploration.

\subsection{Wearable and Touchscreen-Based Vibrotactile Interaction}
Finger-worn and wearable vibrotactile devices provide localized tactile feedback across interaction scenarios such as augmented reality, mixed reality, and touchscreen input (e.g., \cite{vega2024varitouch,asano2014toward,asano2012vibrotactile,chen2018research, yoshimoto2014roughness,terenti2025wear+,cai2024vibopneumo}). These systems can enrich interaction while preserving visual engagement and free hand movement. In particular, placing actuators near the fingertip while keeping the fingerpad available for surface contact makes wearable feedback suitable for touchscreen-based exploration. However, many prior systems are evaluated as device-level augmentations or as techniques for enhancing specific tactile sensations. Less attention has been paid to how users interpret distributed vibrotactile feedback during continuous image exploration. This is especially important in multi-finger interaction, where users may need to coordinate movement, attention, and comparison across multiple tactile channels. This gap motivates our investigation of how finger-worn vibrotactile feedback can support users in making sense of image-derived spatial variation during active touchscreen exploration.

\section{RelicTouch System Design}
\textit{RelicTouch} is a cross-device fingertip vibrotactile system that couples touchscreen exploration with wearable haptic feedback. The system consists of three components: a touchscreen interface, a host-side processing module, and fingertip-worn linear resonance actuators (LRAs). Following prior finger-worn haptic designs~\cite{Preechayasomboon2021}, the actuators were mounted on the dorsal side of the fingers to keep the fingerpad available for direct surface contact. As users slide their fingers across an image, \textit{RelicTouch} modulates vibration intensity in real time according to image-derived spatial features. More prominent or edge-rich regions are mapped to stronger vibration, while flatter or less distinctive regions produce weaker feedback. In this way, tactile cues remain movement-contingent and spatially linked to the displayed image (Figure~\ref{fig:system4}).

\begin{figure}[b]
    \vspace{-12pt}
    \centering
    \includegraphics[width=\linewidth]{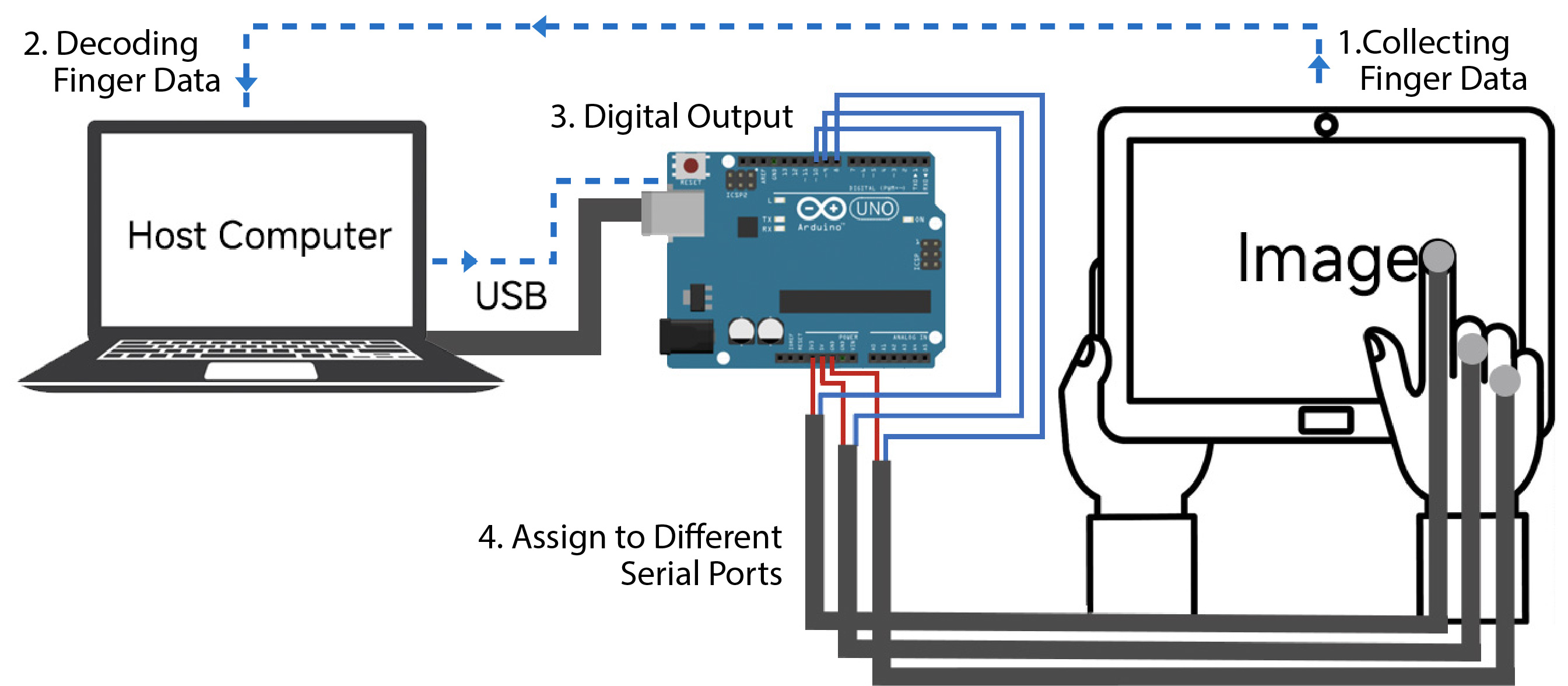}
    \caption{Data flow diagram}
    
    \label{fig:system}
    \vspace{-12pt}
\end{figure}

The prototype was implemented using a custom Android application, a host computer, an Arduino Uno, and three 10$\times$10$\times$4~mm LRAs. The application ran on a 10.1-inch Android tablet and captured touch positions during image exploration. Touch events were sent to the host computer, which converted each contact point into a text-based command containing an actuator ID and a Pulse Width Modulation (PWM) intensity value. These commands were transmitted to the Arduino via serial communication and dispatched to the corresponding LRA. For multi-touch input, each touch point was assigned to an independent haptic channel, allowing multiple actuators to provide concurrent feedback (Figure~\ref{fig:system}).

\begin{figure}[b]
    \vspace{-12pt}
    \centering
    \includegraphics[width=\linewidth]{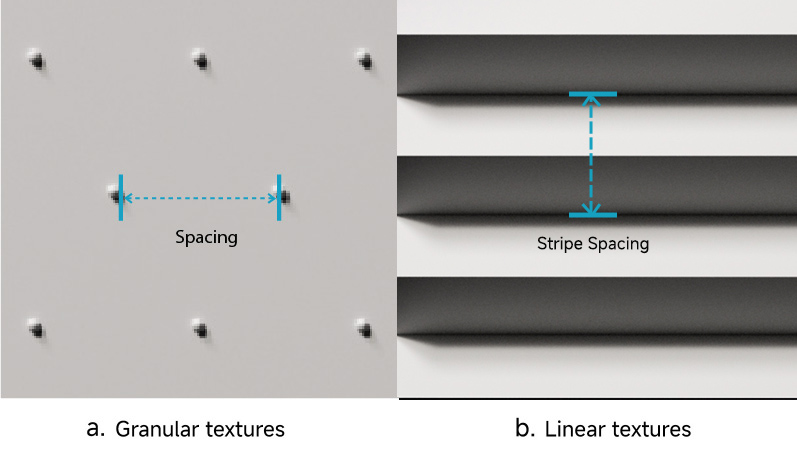}
    \vspace{-18pt}
    \caption{(a).Granular textures consist of granular patterns composed of discrete, spherical elements distributed across the surface, (b).Linear textures: uses a continuous and directional spatial structure}
    \label{fig:system6}
\end{figure}

\begin{figure}[t]
\centering
\includegraphics[width=0.8\columnwidth]{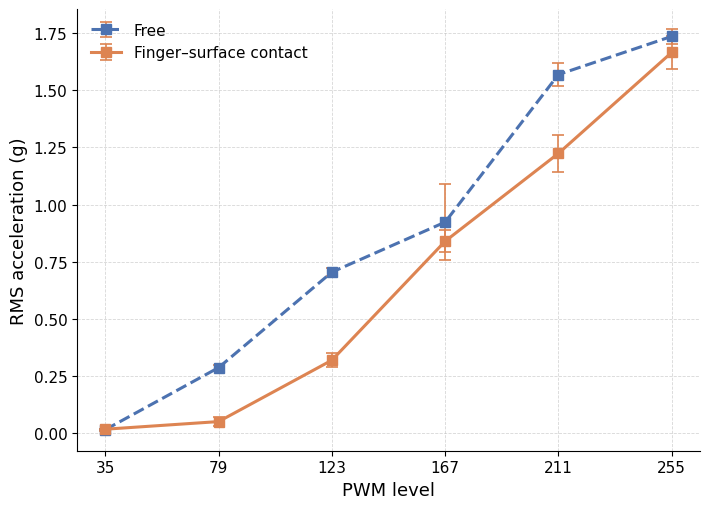}
\vspace{-12pt}
\caption{Actuator calibration across six PWM levels. RMS acceleration was measured under free and finger--surface contact conditions. Both conditions showed an increasing trend with PWM, while contact altered the absolute output magnitude.}
\label{fig:calibration}
\vspace{-15pt}
\end{figure}

As PWM values do not directly correspond to physical vibration amplitude, we conducted an actuator calibration before the user studies. A Micro-Electro-Mechanical System (MEMS) accelerometer (MPU6050) was attached to the LRA housing to record vibration output at the six PWM levels used in Study 1 (see Table~\ref{tab:pwm-stripe-spacing}). For each level, acceleration was measured under two conditions: free vibration and finger-contact vibration, where the actuator was worn on the fingertip during light contact with the touchscreen. We computed the root-mean-square (RMS) acceleration over a fixed sampling window to characterize the relative output intensity. The calibration showed that RMS acceleration increased monotonically with PWM input (Figure \ref{fig:calibration}). Finger contact reduced the absolute amplitude, but the relative ordering between PWM levels remained stable. Across six PWM levels, mean RMS acceleration rose from approximately 0.04 to 1.75~g in free vibration and from 0.04 to 1.63~g under finger--surface contact (Figure~4). Contact changed the absolute magnitude but preserved the increasing relationship between PWM input and measured acceleration. Error bars show the standard deviation across three measurements per level.


\section{Study 1: Formative Parameter-Space Study} \label{study_1}
Study 1 was designed as a formative parameter-space study to establish the perceptual operating conditions for RelicTouch. It consisted of two parts: a preliminary threshold test and a main within-subject experiment. This study used controlled texture stimuli to examine how vibration intensity, stripe spacing, and finger configuration affected haptic-visual congruence and subjective clarity. The results help guide the intensity range and interaction condition of Study 2.

\subsection{Study Design}
The preliminary threshold test used granular dot textures (Figure~\ref{fig:system6}(a)) to determine the usable stimulus range. The stimulus images were created as 1200 $\times$ 800 px images, and pseudo-height units corresponded to pixel-based distances in the generated stimulus images. Thus, we recruited 12 participants for this test and identified 35 PWM as the minimum perceivable vibration level, 25 pseudo-height units as the minimum perceivable dot spacing, and 15 pseudo-height units as the approximate lower bound for distinguishing spacing differences. Based on these values, the main experiment used linear stripe textures (Figure~\ref{fig:system6}(b)) with six vibration-intensity levels and three stripe-spacing values at each level (Table~\ref{tab:pwm-stripe-spacing}). The spacing values indicate the distance between adjacent stripe features. 
In the main experiment, we also included finger configuration as an exploratory design variable. Considering that users may naturally use multiple fingers to compare spatially separated regions on larger touchscreens or exhibition displays, we tested single-finger, two-finger, and three-finger conditions to examine whether additional tactile channels improved haptic-visual congruence or introduced integration costs. For this experiment, we recruited 20 participants (13 female, 7 male), with partial overlap with the preliminary threshold test. The threshold test only determined the stimulus range, and all participants completed training trials before the main experiment to standardize task familiarization.

\begin{table}[h]
\vspace{-10pt}
\centering
\caption{Mapping from vibration intensity (PWM) to pseudo-height values (stripe spacing).}
\label{tab:pwm-stripe-spacing}
\begin{tabular}{cccc}
\toprule
\textbf{PWM} &
\textbf{Spacing 1} &
\textbf{Spacing 2} &
\textbf{Spacing 3} \\
\midrule
255 & 155 & 170 & 185 \\
211 & 125 & 140 & 155 \\
167 &  95 & 110 & 125 \\
123 &  65 &  80 &  95 \\
 79 &  35 &  50 &  65 \\
 35 &  20 &  35 &  50 \\
\bottomrule
\end{tabular}
\vspace{-15pt}
\end{table}

\subsection{Procedure}
We collected the following data: 1) the stripe-spacing with the highest perceptual congruence between the vibration feedback and the visual texture; 2) the overall haptic feedback experience by questionnaire.
Before the formal trials, all participants completed a short training session to familiarize themselves with the wearable actuators, the touchscreen task, and the range of vibrotactile feedback. They were also shown a reference video demonstrating a steady sliding motion. During the formal trials, however, participants were allowed to adjust their movement speed according to their own exploratory strategy given the study aimed to capture how users perceived and compared vibrotactile cues during active touch.

The formal experiment followed a within-subject design and consisted of six blocks corresponding to the six PWM levels in Table~\ref{tab:pwm-stripe-spacing}. Each block included three stripe-spacing conditions and was completed under three finger configurations: single-finger, two-finger, and three-finger touch. The LRA actuators were attached to the index, middle, and ring fingers. In the two-finger condition, participants used the index and middle fingers. In the three-finger condition, they used the index, middle, and ring fingers. The distance between fingers was not strictly controlled, allowing participants to adopt a comfortable and natural multi-finger posture during exploration.

In each trial, participants explored the displayed stripe texture and verbally reported the image number that they perceived as having the highest haptic--visual congruence between the vibration feedback and the visual texture. Each participant completed 54 trials in total (6 intensity levels $\times$ 3 spacing conditions $\times$ 3 finger configurations). After completing the blocks, participants filled in a questionnaire and participated in a brief 2--3 minute debriefing discussion.

\subsection{Questionnaire Design} \label{questionnaire}
Drawing on prior work \cite{terenti2025distal}, we designed an eight-item questionnaire to assess participants' haptic feedback experience. Participants rated each statement (including reverse-coded items to reduce response bias) on a 7-point Likert scale (1 = Strongly Disagree, 7 = Strongly Agree) for each of the three interaction conditions:

{\footnotesize
\begin{itemize}
    \item \textbf{Q1:} I felt I could quickly find the matching texture image using the vibrotactile feedback.
    \item \textbf{Q2:} Moving my finger(s) more slowly helped me better perceive the differences in the images.
    \item \textbf{Q3:} I found the vibrotactile feedback to be effective for distinguishing between different texture images.
    \item \textbf{Q4:} The vibrotactile feedback gave me a sense of depth difference between the texture images.
    \item \textbf{Q5:} The overall process of searching and matching felt seamless.
    \item \textbf{Q6:} I found it confusing to distinguish between texture images using the vibrotactile feedback. (*Reverse-coded*)
    \item \textbf{Q7:} I felt that stronger vibrations corresponded to deeper textures.
    \item \textbf{Q8:} Although I could feel a texture, the sensation did not seem to match the visual image. (*Reverse-coded*)
\end{itemize}}

\begin{figure}[t]
    \centering
    \includegraphics[width=1\linewidth]{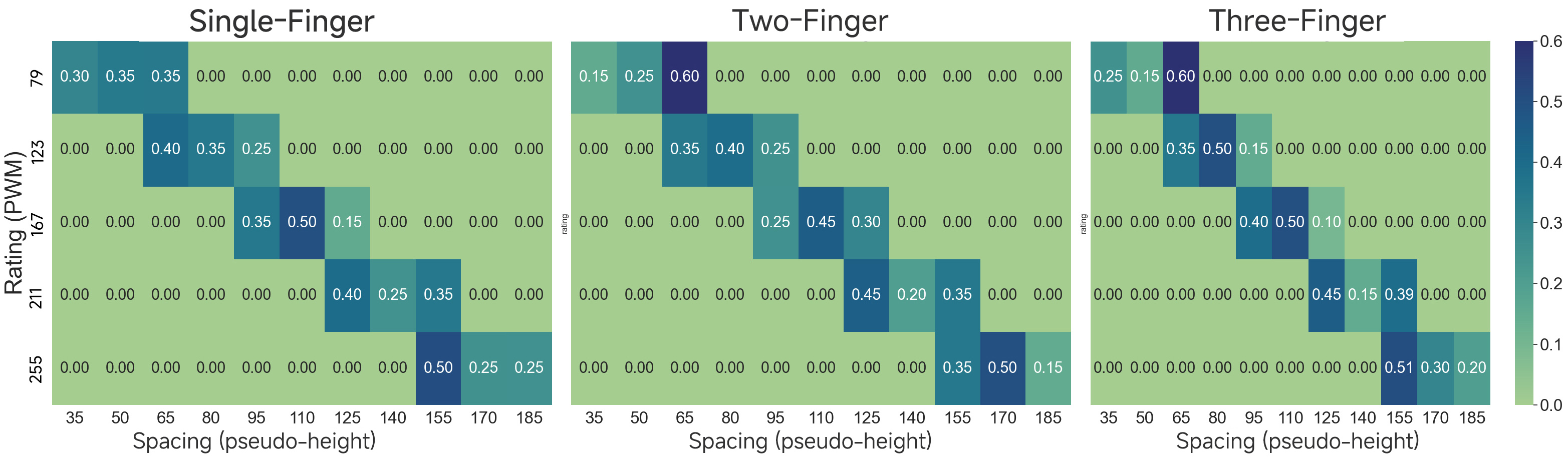}
    \vspace{-20pt}
    \caption{Heatmaps of selection probability across spacing (pseudo-height values) and vibration intensity (PWM) for single-, two-, and three-finger conditions. Each cell represents the Bayesian bootstrap estimate of the probability that a given configuration was selected (binary response = 1), with heavier regions indicating higher preference.}
    \label{fig:Bayesian}
    \vspace{-15pt}
\end{figure}

\subsection{Results}
\textbf{Haptic-Visual Mapping.}
We employed a Bayesian bootstrap approach to estimate selection probabilities without relying on strong parametric assumptions \cite{renganathan2021investigation}. Specifically, Dirichlet-weighted resampling was applied to the observed 0/1 selections to generate a posterior distribution over selection probabilities. The heatmaps (Figure~\ref{fig:Bayesian}) reveal that selections concentrate along a diagonal band consistent with the imposed coupling, but only within restricted regions. This pattern suggests that participants did not accept all imposed intensity–spacing pairings equally. Instead, their selections concentrated in restricted regions, indicating more interpretable subregions within the coupled parameter space. These findings suggest that this reframes the haptic design as identifying perceptually optimal subregions within structured stimulus couplings, rather than defining mappings. Differences across finger configurations further refine this pattern. Multi-finger conditions produce sharper and more concentrated preference zones. For example, both two- and three-finger input show a pronounced peak at 65 pseudo-height units ($\approx0.60$) at 79 PWM, whereas single-finger responses are more distributed ($\approx0.30–0.35$ before 65 pseudo-height units). Similarly, at 123 PWM and 167 PWM, triple-finger input yields narrower peaks (e.g., 80 and 110 pseudo-height units $\approx0.50$), indicating reduced tolerance to suboptimal spacing. In contrast, single-finger input exhibits broader distributions, suggesting weaker spatial selectivity.

\begin{figure}[t]
    \vspace{-12pt}
    \centering
    \includegraphics[width=\linewidth]{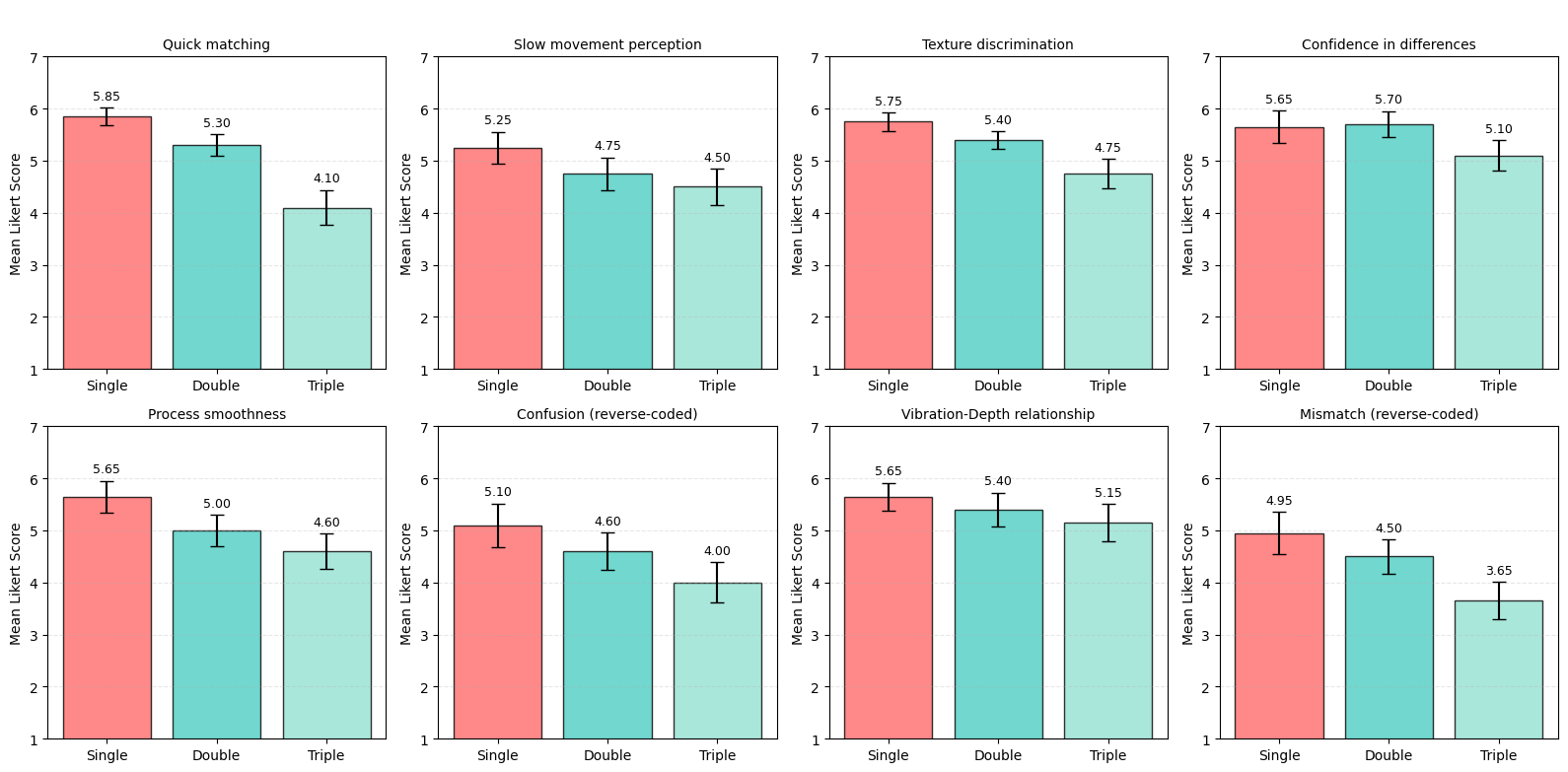}
    \vspace{-20pt}
    \caption{Likert Scale Responses: Comparison Across Finger Configurations}
    \vspace{-20pt}
    \label{fig:system1}
\end{figure} 

\textbf{Finger Number Configurations.}
Responses in Figure~\ref{fig:system1} were analyzed using Friedman tests, given the within-subject design and 7-point Likert-scale measures. Reverse-coded items were recoded before analysis. The aggregated subjective experience score differed significantly across finger configurations, $\chi^2(2)=19.18$, $p<.001$, Kendall's $W=.48$. Ratings were highest for single-finger interaction ($M=5.48$, $SD=.70$), followed by two-finger ($M=5.08$, $SD=.58$) and three-finger interaction ($M=4.49$, $SD=.88$). Holm-corrected Wilcoxon signed-rank tests showed significant differences between single- and two-finger interaction ($p=.003$, $r=.72$), single- and three-finger interaction ($p=.001$, $r=.78$), and two- and three-finger interaction ($p=.003$, $r=.74$). These results suggest that increasing the number of fingers did not uniformly improve subjective experience; single-finger exploration provided higher perceived clarity and congruence than two- or three-finger exploration. We also examined whether vibration detection thresholds were associated with overall subjective ratings. The correlation was not statistically significant ($r=.32$, $p=.224$), suggesting that baseline tactile sensitivity alone did not explain participants' subjective ratings, as interaction-level factors may also have shaped the perceived clarity of the feedback.

\textbf{Brief Debriefing Findings.}
Debriefing comments showed that tactile clarity depended on movement strategy as well as signal parameters. Participants reported that mid-range intensities, especially 123~PWM, felt more natural and interpretable, and that sliding direction, movement speed, finger posture, and contact angle affected perception. Although single-finger input supported clearer localized control, some participants found multi-finger exploration useful for comparing coarse spatial changes across wider stripe spacings. Together with the sharper preference zones observed in two- and three-finger conditions, these findings suggest that multi-finger feedback has conditional value when spatial features are sufficiently separated, rather than being universally clearer than single-finger exploration.


Overall, Study~1 constrained Study~2 rather than directly validating
image interpretation. The 35 PWM threshold distinguished weak baseline feedback from more salient cues, while the intensity findings informed stronger outputs for local transitions; single-finger input was retained for its highest subjective clarity. This reduced interaction variables, allowing Study~2 to focus on movement-based interpretation of continuous image-derived cues, while multi-finger use is revisited in the Discussion for more widely spaced features.




\section{Study 2: Vibrotactile Image Interpretation} 
Study 2 examined how the operating conditions identified in Study 1 translated to continuous image exploration with relief-like visual content.
\subsection{Study Design}

Unlike the fixed intensity--spacing combinations in Study~1,
Study~2 applied the same content-driven mapping to all images.
The 35 PWM threshold served as a reference for reliably salient
feedback rather than a hard lower bound. Values down to 15 PWM
were used only as a weak baseline in flat regions, while stronger
outputs emphasized visual prominence and local transitions.
The 255 PWM value was an upper cap for brief local responses,
rather than a fixed or sustained output.

At each touched pixel, luminance was computed using BT.709
coefficients and normalized to $L_{\mathrm{norm}}\in[0,1]$.
Inverse luminance provided a simple cue for relative visual
prominence, termed \textit{pseudo-depth}, rather than physical
depth:

\begin{equation}
D=1-L_{\mathrm{norm}}^{P_d}, \qquad
I_d=I_{\min}+D(I_{\max}-I_{\min}).
\end{equation}
A raw $3\times3$ Sobel magnitude $G$ emphasized local visual
transitions:
\begin{equation}
I_e=S_eG.
\end{equation}
The final output was:
\begin{equation}
I=
\begin{cases}
\min(I_d,I_{\mathrm{flat}}), & G<G_{\mathrm{thresh}},\\
\min(I_d+I_e,I_{\max}), & G\geq G_{\mathrm{thresh}}.
\end{cases}
\end{equation}
We used $P_d=1.5$, $S_e=0.35$, $G_{\mathrm{thresh}}=80$, $I_{\mathrm{flat}}=35$, $I_{\min}=15$, and $I_{\max}=255$. These values were fixed across all images, with no image-specific tuning. Outputs were
integer-converted and clamped to $[15,255]$ PWM.

\subsection{Participants and Procedure}
Study 2 was intended as an exploratory, expert-informed qualitative evaluation rather than a statistically powered user study. We purposively recruited eight participants (5 female, 3 male) with relevant backgrounds in materials science, computing, electronics, museum experience and interactive design, as these participants were able to reflect on both the experiential qualities and technical limitations of the prototype. The aim was to elicit situated interpretations and design considerations rather than estimate population-level effects.

\begin{figure}[!ht]
    \vspace{-12pt}
    \centering
    \includegraphics[width=\linewidth]{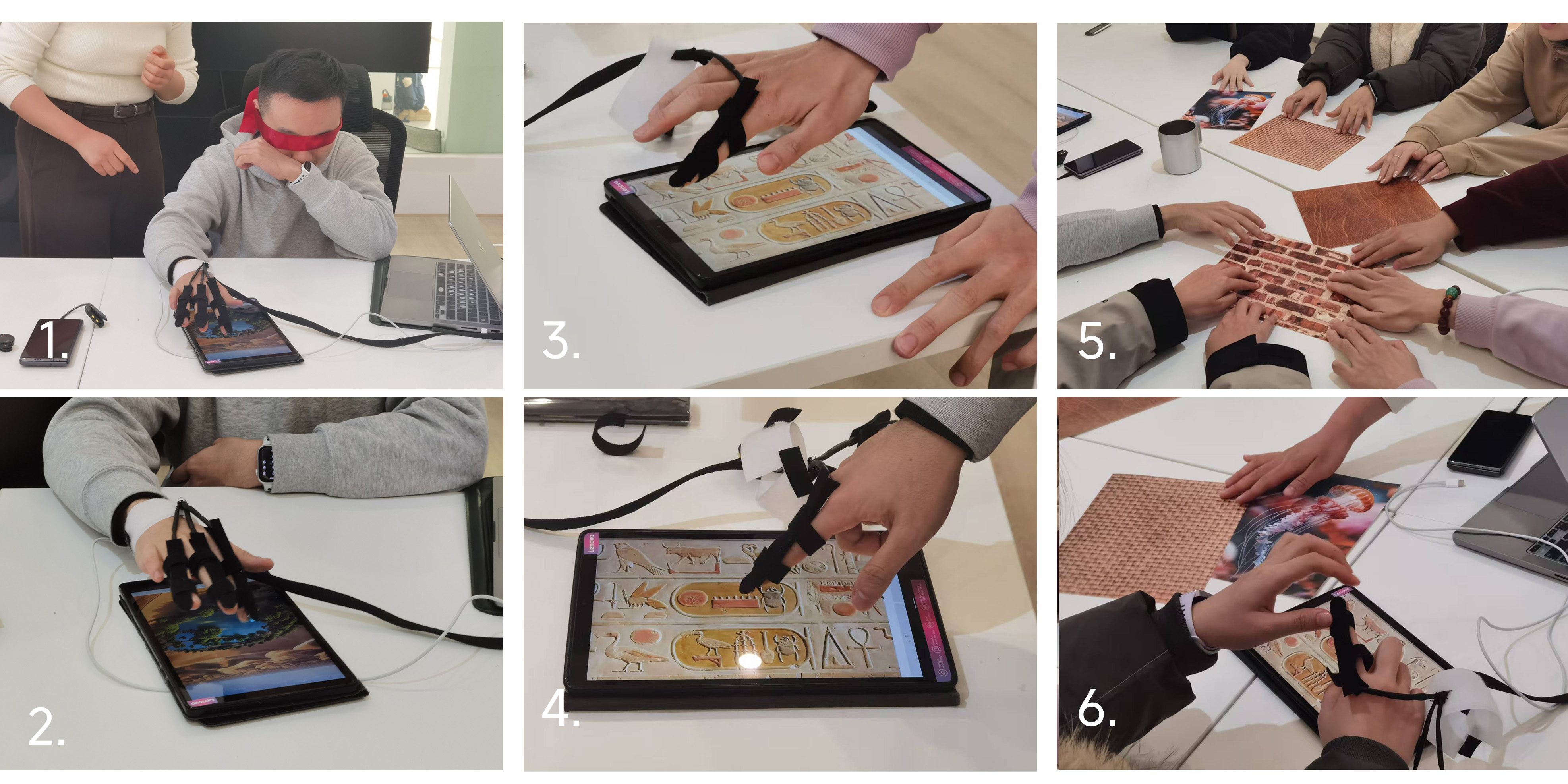}
    \vspace{-17pt}
    \caption{Three experimental tasks. (1–2) Blind exploration: a participant located salient regions in a tablet-displayed image using vibrotactile cues. (3–4) Exploration of an engraved image using RelicTouch. (5–6) Comparison between printed texture samples and RelicTouch-based image exploration.}
    \label{fig:system3}
    \vspace{-15pt}
\end{figure}

To elicit whether fingertip haptics might support more embodied and tangible interactions, this experiment consisted of three activities (see Figure~\ref{fig:system3}): 1) \textbf{Blind haptic description}. A self-selected participant used \textit{RelicTouch} to explore a desert oasis image without vision, verbally describing perceived structures and attempting to locate the oasis based on vibrotactile cues. The participant then re-explored the image with added visual feedback and reflected on changes in interpretation. 2) \textbf{Guided image exploration.} All participants explored one predefined image (e.g., a relief wall carved from Egypt), displayed at 1200$\times$800 pixels, and described how salient features such as lines, patterns and regions were conveyed through vibration. 3) \textbf{Experience sharing and design critique.} Participants discussed the prototype's feasibility, interpretability, and possible use in restricted-touch settings and environments (e.g., museums). 

The group session lasted approximately 60 minutes. Recordings were automatically transcribed, manually checked against the audio, and analyzed using affinity diagramming \cite{lucero2015using}. Two authors independently reviewed the transcripts, developed initial notes, and collaboratively grouped recurring observations into three themes; disagreements were resolved through discussion.

\subsection{Results and Findings}

\textbf{Legible Encoding Requires Framing.}
Participants interpreted vibrotactile feedback less as a simulation of material texture and more as an encoding of image structure. They focused on whether vibration changes were distinct, consistent, and useful for identifying salient regions. P6 described the experience as \textit{``image information was encoded and decoded through touch,''} suggesting that spatial meaning was inferred through interpretation rather than physical realism. In the blind exploration task, P3 located the oasis region based only on vibrotactile variation and described the feedback as \textit{``clear and reliable.''} However, participants also noted that cues became easier to interpret when supported by task context, visual reference, or prior expectation. Several participants (P2, P3, P5) found fine details difficult to resolve on a tablet-sized display, while others (P4, P5) suggested that larger displays could make spatial variation more legible. These comments suggest that vibrotactile legibility depends not only on signal design, but also on framing, display scale, and spatial resolution.


\textbf{Movement-Based Depth and Shape Perception.} Participants described depth perception as emerging through movement rather than static contact. Stronger vibration was generally associated with more prominent or salient regions, while weaker feedback indicated flatter or less distinctive areas. As P2 noted, interpretation involved \textit{``feeling changes as the finger moves,''} emphasizing comparison across space rather than isolated tactile events. However, participants also reported limits in shape specificity. While relative prominence and local contrast were easier to perceive, precise geometric forms were harder to identify. For example, P2 noted that although a triangular pattern was known to exist, the system mainly conveyed intensity variation rather than an explicit contour. These comments suggest that participants interpreted the feedback primarily as changing spatial prominence, rather than as a direct representation of exact shape.

\textbf{Cross-Modal Scaffolding in Situated Use.}
Participants positioned RelicTouch as a scaffold for interpreting visual content, rather than as a replacement for vision or physical touch. When discussing possible situated uses, including museum and cultural heritage contexts, several participants (P1, P2, P3, P5) suggested that vibrotactile feedback could guide attention to localized details and structurally meaningful regions in visually accessible but untouchable artifacts. However, participants also noted scope limitations: P6 and P7 suggested that full 3D objects would be difficult to understand through vibration alone, making the approach more suitable for fragments, relief-like images, or regions with clear spatial variation. These comments indicate that vibrotactile cues became more intelligible when paired with visual references, task goals, or curatorial framing, supporting their role as an interpretive layer within situated visual interaction.

\section{Discussion and Design Implications}
Based on our research findings, We derive three situated design implications for fingertip-based vibrotactile image exploration on touchscreens.

\textbf{Prioritize Separable Vibrotactile Encoding.}
Across both studies, vibrotactile feedback functioned less as a direct simulation of material texture and more as an interpretable encoding of image structure. In Study 1, the Bayesian heatmaps (Figure~\ref{fig:Bayesian}) showed that participants' selections concentrated in restricted regions of the intensity--spacing parameter space, suggesting that vibrotactile cues became legible only when they formed distinguishable perceptual patterns. Study 2 supported this interpretation, as participants relied on consistent vibration changes to identify salient structures and described the experience as decoding spatial information through touch. We refer to this requirement as signal separability: the extent to which vibrotactile variations form stable and distinguishable contrasts during exploration. Designers should therefore emphasize boundaries and meaningful spatial transitions rather than fine-grained material realism.

\textbf{Support Movement-Contingent Exploration.}
Prior work has described tactile perception as active and motion-coupled, emerging through exploratory interaction with physical objects~\cite{aliabbasi2025haptic}. Our findings build on this view in designed vibrotactile systems, where tactile information is not physically present on the surface but encoded through vibration and interpreted through movement. We use \textit{movement-contingent} to emphasize how users make sense of changing signals over time, shifting attention from system-level coupling to active spatial interpretation. Across the two studies, participants compared changes in vibration along movement trajectories to infer relative spatial prominence. This mechanism supported the perception of gradients, salience, and structural variation, but was less effective for precise geometric contours. Designers should therefore prioritize progression, contrast, and exploratory dynamics over static point-by-point precision, and combine vibrotactile cues with guided scanning or visual anchors when more specific shape interpretation is required.


\textbf{Match Feedback Channels to Image Scale.}
Study~1 showed higher subjective ratings for single-finger
exploration but narrower multi-finger preference zones in some
parameter regions. Although cognitive load was not measured
directly, this pattern suggests a serial--parallel trade-off.
Single-finger exploration preserves a direct correspondence
between one touch location and one signal, reducing divided
attention and supporting fine-grained localization, but it
requires serial scanning. Multi-finger exploration can sample
wider regions concurrently and may aid comparison of separated
coarse features, yet users need to coordinate finger positions
and integrate simultaneous signals affected by posture and
contact. In densely patterned or high-complexity images, these
integration demands may offset the benefit of parallel coverage.
Designers should therefore match channel number to feature
spacing and task complexity rather than assume that more
channels provide richer perception.

\section{Conclusion}

In this work, we present \textit{RelicTouch}, a fingertip-worn vibrotactile system that encodes image-derived spatial variation as movement-contingent tactile feedback for touchscreen exploration. Through a formative parameter-space study and an expert-informed qualitative exploration, we examined how intensity, spatial variation, finger configuration, and movement shape users' interpretation of image-based tactile cues. Our findings suggest that interpretable vibrotactile feedback is associated with signal separability, exploratory movement, visual framing, and image scale. Rather than simulating material properties directly, \textit{RelicTouch} shows how wearable vibrotactile feedback can support image interpretation as an active, situated process.

\bibliographystyle{IEEEtran} 
\bibliography{reference}

\end{document}